\documentclass[sigconf]{acmart}

\AtBeginDocument{%
  \providecommand\BibTeX{{%
    \normalfont B\kern-0.5em{\scshape i\kern-0.25em b}\kern-0.8em\TeX}}}

\setcopyright{rightsretained}
\copyrightyear{2021}
\acmYear{2021}
\acmDOI{10.1145/xxxxxxx.xxxxxxx}

\acmConference[arXiv '21]{arXiv '21}{Feb 07, 2021}{Online}
\acmBooktitle{arXiv '21, Feb 07, 2021, Online}
\acmPrice{15.00}
\acmISBN{978-x-xxxx-xxxx-x/YY/MM}

\usepackage{graphicx}
\usepackage{amsmath}
\usepackage{multirow}
\usepackage{booktabs}
\usepackage{array}
\usepackage{subfigure}
\usepackage{amssymb}
\usepackage{algorithm}
\usepackage{verbatim}
\usepackage{bm}     
\usepackage{enumitem}       

\usepackage{algpseudocode}

\begin{document}

\fancyhead{}

\title{Improving Accuracy and Diversity in Matching of Recommendation with Diversified Preference Network}


\author{Ruobing Xie}
\authornote{Both authors contributed equally to this research.}
\affiliation{\institution{WeChat, Tencent}
\city{Beijing}
\country{China}}
\email{ruobingxie@tencent.com}

\author{Qi Liu}
\authornotemark[1]
\affiliation{\institution{WeChat, Tencent}
\city{Beijing}
\country{China}}
\email{addisliu@tencent.com}

\author{Shukai Liu}
\affiliation{\institution{WeChat, Tencent}
\city{Beijing}
\country{China}}
\email{shukailiu@tencent.com}

\author{Ziwei Zhang}
\affiliation{\institution{Tsinghua University}
\city{Beijing}
\country{China}}
\email{zw-zhang16@mails.tsinghua.edu.cn}

\author{Peng Cui}
\affiliation{\institution{Tsinghua University}
\city{Beijing}
\country{China}}
\email{cuip@tsinghua.edu.cn}

\author{Bo Zhang}
\affiliation{\institution{WeChat, Tencent}
\city{Beijing}
\country{China}}
\email{nevinzhang@tencent.com}

\author{Leyu Lin}
\affiliation{\institution{WeChat, Tencent}
\city{Beijing}
\country{China}}
\email{goshawklin@tencent.com}


\begin{abstract}
  Recently, real-world recommendation systems need to deal with millions of candidates. It is extremely challenging to conduct sophisticated end-to-end algorithms on the entire corpus due to the tremendous computation costs. Therefore, conventional recommendation systems usually contain two modules. The matching module focuses on the coverage, which aims to efficiently retrieve hundreds of items from large corpora, while the ranking module generates specific ranks for these items. Recommendation diversity is an essential factor that impacts user experience. Most efforts have explored recommendation diversity in ranking, while the matching module should take more responsibility for diversity. In this paper, we propose a novel Heterogeneous graph neural network framework for diversified recommendation (GraphDR) in matching to improve both recommendation accuracy and diversity. Specifically, GraphDR builds a huge heterogeneous preference network to record different types of user preferences, and conduct a field-level heterogeneous graph attention network for node aggregation. We also innovatively conduct a neighbor-similarity based loss to balance both recommendation accuracy and diversity for the diversified matching task. In experiments, we conduct extensive online and offline evaluations on a real-world recommendation system with various accuracy and diversity metrics and achieve significant improvements. We also conduct model analyses and case study for a better understanding of our model. Moreover, GraphDR has been deployed on a well-known recommendation system, which affects millions of users. The source code will be released.
\end{abstract}




\maketitle

\section{Introduction}

\begin{figure}[!hbtp]
\centering
\includegraphics[width=0.96\columnwidth]{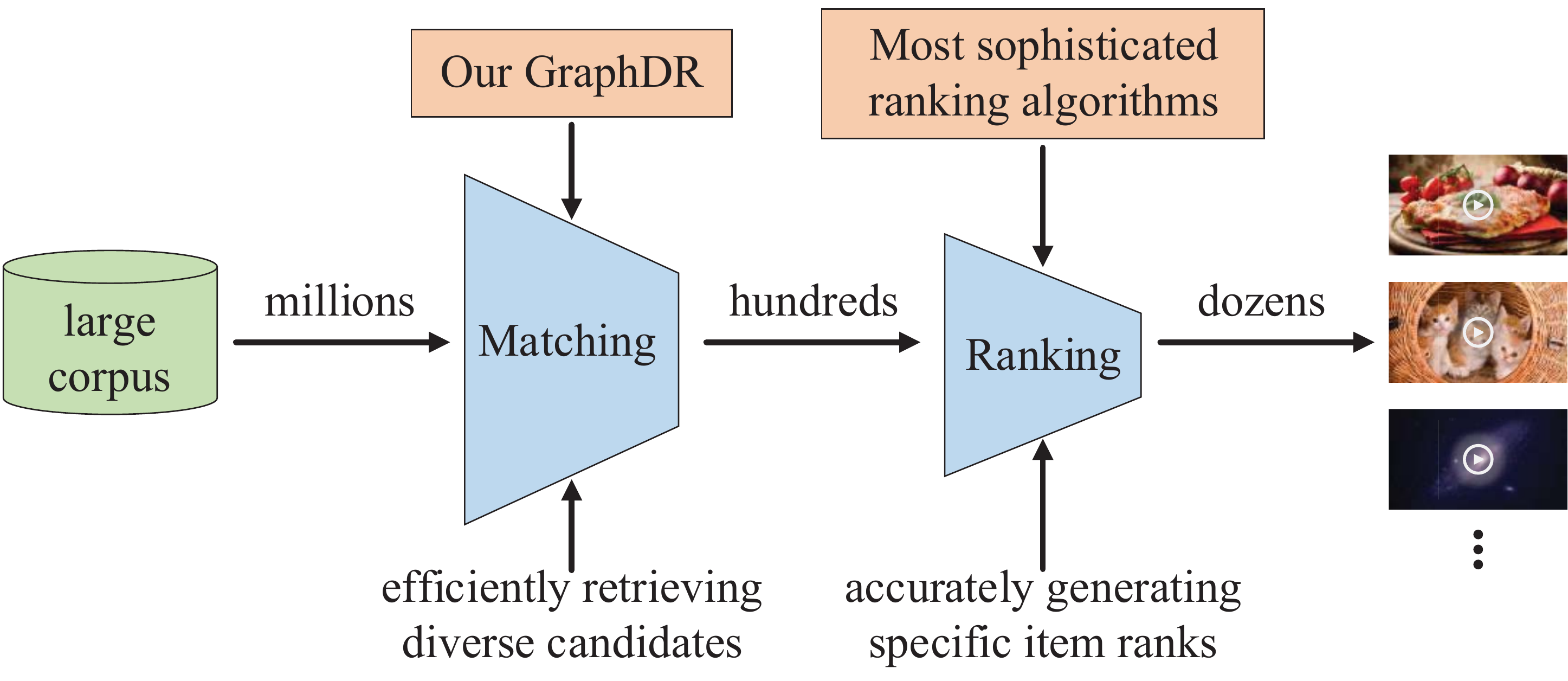}
\caption{An example of a real-world recommendation system. GraphDR focuses on the matching module, which aims to retrieve user-interested and diverse items efficiently.}
\label{fig:example}
\end{figure}

Recently, real-world personalized recommendation systems usually need to deal with hundreds of millions of items \cite{wang2018billion}. Therefore, it is challenging to conduct complicated end-to-end recommendation algorithms on the entire corpus, for even a linear time complexity
w.r.t the corpus size is unacceptable \cite{zhu2018learning}. To balance both effectiveness and efficiency in real-world scenarios, conventional recommendation systems usually consist of two modules, namely the matching module and the ranking module \cite{covington2016deep,xie2020internal}. The \textbf{matching} module, also regarded as the candidate generation in the Youtube model \cite{covington2016deep}, aims to retrieve a small subset of (usually hundreds of) items from the entire corpus efficiently. In contrast, the \textbf{ranking} module conducts sophisticated models on these retrieved items to get specific item ranks. Fig. \ref{fig:example} shows the classical two-step architecture. The matching module concentrates more on the diversity, efficiency and item coverage, while the ranking module focuses more on the accuracy of specific item ranks. This two-step architecture balances efficiency and effectiveness in practice.

Conventional recommendation models usually regard recommendation accuracy metrics such like Click-through-rate (CTR) as their central objectives, in which popular items clicked by users are more preferred.
However, such objectives will lead to homogenization issues that reduce personalization and harm user experiences.
To solve this issue, \textbf{recommendation diversity} is considered to evaluate the overall recommendation performances from another aspect \cite{bradley2001improving}.
It is measured in two classical ways: the individual diversity and the aggregate diversity \cite{kunaver2017diversity}. The \textbf{individual diversity} focuses on the local diversity in each recommended item list, which aims to balance user-item similarities and item-item dissimilarities \cite{chen2018fast}. In contrast, the \textbf{aggregate diversity} focuses on the global diversity in the overall recommendation, which is usually measured by the coverage of items that could be recommended by models in the entire corpus \cite{karakaya2018effective}.
The significance of diversity has been widely verified to provide highly idiosyncratic items in recommendation \cite{zhang2019empirical}, which should be considered in real-world scenarios.

There are lots of ranking models that have explored recommendation diversities with the help of dissimilarity factors \cite{bradley2001improving}, external taxonomy information \cite{ziegler2005improving}, clustering \cite{aytekin2014clustering} and graphic technologies \cite{nandanwar2018fusing}.
However, most diversified recommendation models are specially designed for ranking, which are incredibly time-consuming to be used in matching with millions of items \cite{nandanwar2018fusing}, while very few works systematically focus on the diversity in matching. In fact, matching should take more responsibility for diversity, since it cares more about the coverage of user-interested items rather than their specific item ranks. The recommendation diversity needs to be first guaranteed in the matching module. Otherwise, the homogenization of the item candidates generated by the matching module will inevitably lead to the lack of diversity in the final recommendation.

In this paper, we aim to improve both recommendation accuracy and diversity in the matching module, which is essential in real-world recommendation systems. We propose a novel \textbf{Heterogeneous graph neural network framework for diversified recommendation (GraphDR)}. Precisely, GraphDR mainly consists of three modules:
(1) \emph{Diversified preference network construction}, which aims to build a huge global heterogeneous network containing various interactions between different types of nodes including videos, tags, medias, users and words. These interactions between essential recommendation factors reflect user diverse preferences from a global view, which are the sources of diversity.
(2) \emph{Heterogeneous network representation learning (NRL)}, which learns node representations with a novel field-level heterogeneous graph attention network (FH-GAT). FH-GAT helps to better maintain and aggregate different types of interactions.
We also innovatively conduct a neighbor-similarity based objective to encode user diverse preferences into heterogeneous node representations. Different from CTR-oriented objectives that simply focus on click behaviors, the neighbor-similarity based objective highlights diversity by considering multiple factors of videos such as user watching habit, audience community, video content, video taxonomy, and content provider.
(3) \emph{Online multi-channel matching}, which generates a small subset of user-interested and diverse item candidates efficiently through multiple channels.
The multi-channel strategy is conducted to further amplify the diversity in the final results.
The diversity derives from all three modules in GraphDR.

In experiments, we conduct both offline and online evaluations on a real-world video recommendation system, which is widely used by hundreds of millions of users.
We conduct extensive experiments to measure the recommendation accuracy and diversity with dozens of metrics. We also explore GraphDR with model analyses, ablation tests and case studies for better understanding. The main contributions are concluded as follows:
\begin{itemize}
\item We highlight and systematically explore the recommendation diversity issue in the matching module, which is essential in practical large-scale recommendation systems.
\item We propose a novel GraphDR framework to jointly improve both recommendation accuracy and diversity in real-world matching. To the best of our knowledge, we are the first to introduce GNN on heterogeneous preference networks for diversified recommendation in matching.
\item We propose a novel field-level heterogeneous GAT model to aggregate neighbors with different feature fields. We also innovatively conduct the neighbor-similarity based loss to polish recommendation diversity.
\item The offline and online evaluations indicate that GraphDR can improve both accuracy and diversity in practice. GraphDR is simple and effective, which has been deployed on a real-world recommendation system used by millions of users. It is also convenient to adopt GraphDR to other scenarios.
\end{itemize}

\section{Related Works}

In related works, we first give a brief introduction to the classical recommendation algorithms, and then introduce the efforts in recommendation diversity. We also include a discussion on the graph neural networks used in recommendation.

\subsection{Recommendation Systems}

Collaborative filtering (CF) is a classical method which recommends items with similar items or users \cite{sarwar2001item}.
Matrix factorization (MF) attempts to decompose user-item interaction matrix to get user and item representations \cite{koren2009matrix}.
FM \cite{rendle2010factorization} expands to model second-order feature interactions with latent vectors.
However, most neural ranking models rely on user-item interactions for prediction. Hence, these complicated ranking models are hard to be directly used in matching, for they are extremely time-consuming when handling million-level items.
With the thriving in deep learning, neural models like Deep Crossing \cite{shan2016deep}, FNN \cite{zhang2016deep}, PNN \cite{qu2016product}, Wide\&Deep \cite{cheng2016wide}, DCN \cite{wang2017deep} and DFN \cite{xie2020deep} are proposed to improve recommendation performances. DeepFM \cite{guo2017deepfm}, AFM \cite{xiao2017attentional} and NFM \cite{he2017neural} improve the original FM with DNN or attention.
AutoInt \cite{song2019autoint} and BERT4Rec \cite{sun2019bert4rec} also brings in self attention. Recently, AFN \cite{cheng2020adaptive} and AutoFIS \cite{liu2020autofis} are proposed to smartly model high-order feature interactions via logarithmic transformation or automatic feature selection.
Most deep ranking models are challenging to be utilized in real-world matching module, for they are extremely time-consuming dealing with millions of candidates.

In contrast, there are much fewer works specially designed for matching. Conventional systems usually use IR-based methods \cite{khribi2008automatic} or Collaborative filtering (CF) based methods \cite{sarwar2001item} for fast retrieval. For neural models, embedding-based retrieval such as DSSM \cite{huang2013learning} is also widely deployed.
Recently, Youtube \cite{covington2016deep} brings in deep models to learn user preference in matching. Moreover, TDM \cite{zhu2018learning}, JTM \cite{zhu2019joint} and OTM \cite{zhuo2020learning} arrange items with tree structures to accelerate top-n item retrieval, which combine matching and ranking in a single model. ICAN \cite{xie2020internal} is specially designed for cold-start multi-channel matching. \citeauthor{huang2020embedding} \shortcite{huang2020embedding} also proposes an industrial embedding-based retrieval framework in Facebook search.
However, these matching models mainly focus on CTR-oriented objectives. It is still challenging for these models to balance accuracy and diversity in real-world scenarios. In this work, we aim to improve both recommendation accuracy and diversity in the matching module via the proposed GraphDR with the diversified preference network.

\subsection{Diversified Recommendation}

Merely using CTR-oriented objectives will make hot items hotter, which inevitably brings in serious homogenization issues that may degrade user experiences \cite{zhang2008avoiding}.
The significance of diversity has been verified by lots of efforts, since it could provide highly idiosyncratic items with less homogeneity for users in personalized recommendation \cite{bradley2001improving,zhang2019empirical}.
Recommendation diversity is mainly measured in individual diversity and aggregate diversity \cite{kunaver2017diversity}.
The individual diversity focuses on the local diversity in recommended list. \cite{bradley2001improving} and \cite{ziegler2005improving} focus on intra-list item dissimilarities. \cite{zhang2008avoiding} proposes a novel item novelty, which measures the additional information from new items. Some works measure diversity with the varieties of taxonomy in item lists \cite{ziegler2005improving}.
In contrast, the aggregate diversity measures the global diversity in overall systems. \cite{karakaya2018effective} measures this diversity with the coverage of recommended items.
The higher item coverage indicates that the model could recommend more long-tail items, which implies a more diversified system from the global aspect.

There are some works that model diversity in ranking. \cite{bradley2001improving} bring dissimilarity factors to the loss functions to measure the individual diversity.
External taxonomy information (e.g., tag, category and subtopic) \cite{ziegler2005improving,wu2016relevance} and knowledge graph \cite{gan2020enhancing} are useful factors for diversity.
Other technologies such as entropy regularizer \cite{qin2013promoting}, clustering \cite{aytekin2014clustering}, graph-based models \cite{zhu2007improving,mei2010divrank,nandanwar2018fusing}, and greedy map inference \cite{chen2018fast} have also been explored for diversified recommendation.
Recently, diversified recommendation is armed with reinforcement learning \cite{liu2019diversity} and adversarial learning \cite{wu2019pd}. Recommendation bandits \cite{li2016collaborative,mahadik2020fast} are also well explored.
However, most diversified models are specially designed for ranking, which are hard to be directly used in matching. To the best of our knowledge, we are the first to use GNN on the global heterogeneous interactions to improve both accuracy and diversity in the matching module.

\subsection{Graph Neural Network (GNN)}

Recently, GNN has been widely explored and verified in various fields. GCN \cite{kipf2017semi} introduces convolution to graphs based on spectral graph theory.
GraphSAGE \cite{hamilton2017inductive} conducts an inductive representation learning on large graphs. Graph attention network (GAT) \cite{velivckovic2018graph} brings in graph attention mechanism. HetGNN \cite{zhang2019heterogeneous} and HAN \cite{wang2019heterogeneous} extend GNN to heterogeneous networks.
In recommendation, \citet{wu2019session}, \citet{fan2019graph} and \citet{he2020lightgcn} further use GNN on session-based and social-based recommendation. Heterogeneous graphs are also widely adopted to model different types of essential objects such as users, items, tags and providers in recommendation \cite{lu2020social,liu2020graph}.
Inspired by these models, we conduct a heterogeneous graph to model various types of feature interactions, and also use a heterogeneous GNN model for node aggregation.

\begin{figure*}[!htbp]
\centering
\includegraphics[width=0.95\textwidth]{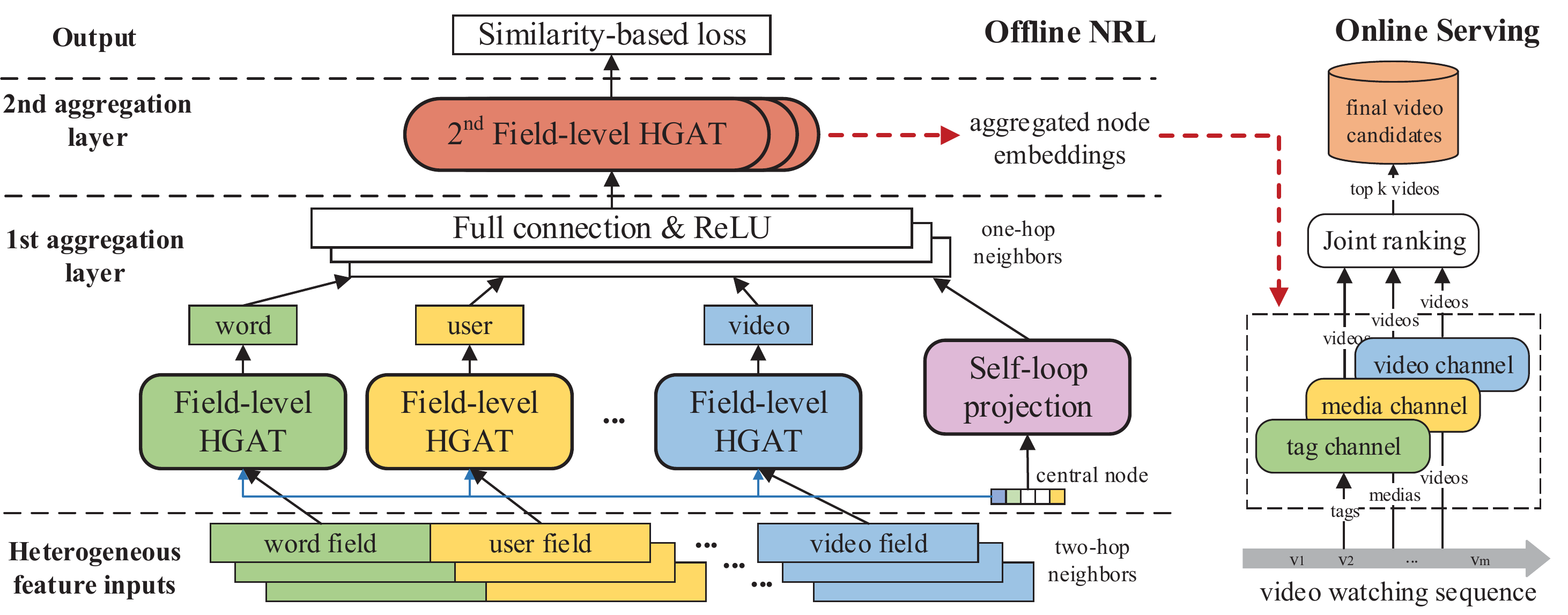}
\caption{The offline NRL and online serving parts of GraphDR for matching in recommendation. The left offline NRL part is the proposed FH-GAT model, which builds the aggregated node embeddings with heterogeneous GAT on the diversified preference network. The right online multi-channel matching part aims to retrieve hundreds of videos from large corpora efficiently. The recommendation diversity comes from diversified preference network, FH-GAT trained with diversity-enhanced training objective, and online multi-channel matching.}
\label{fig:architecture}
\end{figure*}

\section{Methodology}

In this paper, we propose GraphDR to improve both accuracy and diversity in matching by considering user diverse preferences. In this section, we first show the overall framework of GraphDR (Sec. \ref{sec.framework}). Second, we introduce the construction of nodes and edges in the diversified preference network, which is the source of diversity in our model (Sec. \ref{sec.diversified_network}). Next, we introduce the Diversity-enhanced network representation learning model FH-GAT used to generate node representations for all types of nodes (Sec. \ref{sec.NRL}). Finally, we give a detailed discussion on the proposed Diversity-enhanced training objective (Sec. \ref{sec.objective}). We further introduce the online deployment of the multi-channel matching module (Sec. \ref{sec:online_serving}).

\subsection{Overall Architecture}
\label{sec.framework}

The GraphDR framework mainly contains three modules as in Fig \ref{fig:architecture}, including diversified preference network construction, network representation learning, and online multi-channel matching. In offline NRL, GraphDR first collects various informative interactions between heterogeneous nodes to build a huge global diversified preference network. Next, we propose a field-level HGAT model to learn node embeddings with the neighbor-similarity based objective. In online serving, the multi-channel matching retrieves hundreds of accurate and diverse item candidates efficiently with multiple channels.
The offline NRL conducts time-consuming training to encode user diverse preferences into node embeddings, while the online serving efficiently uses these learned embeddings for fast and diversified multi-channel retrieval.

\subsection{Diversified Preference Network}
\label{sec.diversified_network}

The diversified preference network is the fundamental of diversity. We attempt to bring in heterogeneous interactions between essential objects in recommendation to describe user diverse preferences.
Precisely, we focus on five different types of nodes including \textbf{video}, \textbf{tag}, \textbf{media}, \textbf{user} and \textbf{word}, which are essential factors that may impact users in recommendation.
Each video has a title (containing words) and several tags annotated by editors. The video provider is viewed as the media. To alleviate the data sparsity and reduce computation costs, we cluster users into user groups as communities according to their basic profiles (i.e., the gender-age-location attribute triplets in this work), and consider these user groups as user nodes. We group users via user basic profiles for higher coverage.

We assume that the interactions between these five types of objects can reflect user diverse preferences. In GraphDR, we consider six types of edges to record these multi-aspect preferences as:
\begin{itemize}
\item \textbf{Video-video edge.} We generate the video-video edge between two video nodes if they have appeared adjacently in a user's video session. To reduce noises, we only use the \textbf{valid watching behaviors}, where videos have been watched for more than $70\%$ of their total time lengths. Video-video edges record the sequential user watching habits in sessions.
\item \textbf{Video-user edge.} Video-user edges are built if a video is validly watched by a user group at least $3$ times in a week. This edge stores coarse-grained user-item interactions and also implies the audience community of videos.
\item \textbf{Video-tag edge.} Video-tag edge connects videos with their corresponding tags, which reflects the coarse-grained semantic preferences of taxonomy in videos.
\item \textbf{Video-word edge.} Video-word edge links videos with their words in titles, which reflects the fine-grained semantic preferences of detailed word-level contents in videos.
\item \textbf{Video-media edge.} Video-media edges are drawn between videos and their medias, which shows the video providers.
\item \textbf{Tag-tag edge.} We build tag-tag edges according to tag co-occurrence in a video, which highlights taxonomy relevance.
\end{itemize}
All edges are undirected and unweighted.
These heterogeneous edges bring in additional information of videos besides user-item click behaviors. They can reflect user diverse preferences in user watching habit, audience community, video content, taxonomy, and content provider.
For instance, two related videos may be linked via the same user groups, video providers, tags or watching sessions, or even connected by a multi-step path containing heterogenous nodes.
The multi-hop paths via heterogeneous nodes and edges build up the potential reasons for recommendation, which are implicit, low-correlational but diversified.
It is also not difficult to extend other types of nodes and edges in GraphDR.

\subsection{Diversity-enhanced Network Representation Learning}
\label{sec.NRL}

Network representation learning aims to encode user diverse preferences into node representations. Inspired by \cite{wang2019heterogeneous,liu2020graph}, we propose a new \textbf{Field-level Heterogeneous Graph Attention Network (FH-GAT)}. Fig. \ref{fig:architecture} shows the 2-layer architecture.

\subsubsection{Heterogeneous Feature Layer}

We first project all heterogeneous nodes into the same feature space. For the k-th node, its overall neighbor set $N_k$ could be divided into five \emph{\textbf{feature fields}} according to their types as $\{\bm{\bar{v}}_k, \bm{\bar{t}}_k, \bm{\bar{m}}_k, \bm{\bar{u}}_k, \bm{\bar{d}}_k\}$, where $\bm{\bar{v}}_k$, $\bm{\bar{t}}_k$, $\bm{\bar{m}}_k$, $\bm{\bar{u}}_k$ and $\bm{\bar{d}}_k$ indicate the one-hot representations of video, tag, media, user, word neighbors respectively.
The node feature embeddings of the k-th node $\bm{h}_k$ is as follows:
\begin{equation}
\begin{split}
\bm{f}_k=\mathrm{concat}(\bm{v}_k,\bm{t}_k,\bm{m}_k,\bm{u}_k,\bm{d}_k),
\end{split}
\end{equation}
in which $\bm{v}_k$ indicates the video-field feature embedding. In this work, we empirically set $\bm{v}_k=\bm{P}_v\bm{\bar{v}}_k$, where $\bm{P}_v \in \mathbb{R}^{d_v\times n_v}$ represents the lookup projection matrix generating $\bm{v}_k$ with the video neighbors. $d_v$ is the dimension of $\bm{v}_k$ and $n_v$ is the number of video nodes. For efficiency, the projection matrix is pre-defined as the indicator of top-frequent video neighbors and fixed during training.
$\mathrm{concat}(\cdot)$ is the concatenation operation.
The tag, media, user and word field feature embeddings $\bm{t}_k$, $\bm{m}_k$, $\bm{u}_k$ and $\bm{d}_k$ are generated similarly as the video field feature embedding $\bm{v}_k$.

\subsubsection{Field-level HGAT Layer}

This layer takes the neighbor feature embeddings $\{\bm{f}_{1}, \cdots, \bm{f}_{l}\}$ of the k-th node as inputs.
We set a weighting vector group $\{\bm{w}_k^v, \bm{w}_k^t, \bm{w}_k^m, \bm{w}_k^u, \bm{w}_k^d\}$ for each field, where $\bm{w}_k^v$ represents the k-th weighting vector of video. The output embedding $\bm{y}_k^v$ of the video field is defined as follows:
\begin{equation}
\begin{split}
\bm{y}_k^v=\sum_{i=1}^{l}\alpha_{ki}^{v}\bm{v}_{i}, \quad \alpha_{ki}^{v}=\frac{\exp({\bm{w}_k^v}^\top\bm{v}_{i})}{\sum_{j=1}^{n}\exp({\bm{w}_k^v}^\top\bm{v}_{j})},
\end{split}
\end{equation}
where $\alpha_{ki}^{v}$ is the weight of the k-th node to its i-th neighbor in the video field. The construction of $\bm{y}_k^t$, $\bm{y}_k^m$, $\bm{y}_k^u$ and $\bm{y}_k^d$ are the same as $\bm{y}_k^v$. We concatenate these embeddings to form the final neighbor-based representation $\bm{y}_k^{N}$ as follows:
\begin{equation}
\begin{split}
\bm{y}_k^{N}=\mathrm{ReLU}(\bm{W}_{n} \cdot \mathrm{concat}(\bm{y}_k^v,\bm{y}_k^t, \bm{y}_k^m,\bm{y}_k^u,\bm{y}_k^d)).
\end{split}
\end{equation}
We further consider the self-loop projection as a supplement to highlight the central k-th node's information. We have:
\begin{equation}
\begin{split}
\bm{y}_k^{S}=\mathrm{ReLU}(\bm{W}_{s} \cdot \bm{f}_k).
\end{split}
\end{equation}
Next, we combine neighbor and self-loop based representations to get the 1st layer output $\bm{y}_k$, and use the 2nd FH-GAT layer to get the final aggregated representation $\bm{h}_k$ as:
\begin{equation}
\begin{split}
\bm{h}_k=\mathrm{FH}\mathrm{-}\mathrm{GAT}(\bm{y}_k), \quad
\bm{y}_k=\lambda_s \cdot \bm{y}_k^{S} + (1-\lambda_s) \cdot \bm{y}_k^{N},
\end{split}
\end{equation}
where $\lambda_s$ is empirically set as $0.5$.

FH-GAT aggregates heterogeneous neighbors separately in each feature field with different field-specific attention, which delicately encodes user diverse preferences related to specific fields to the final node representation.
Other GNN models could also be easily adapted to our framework. Comparing with other heterogeneous GAT models like \cite{wang2019heterogeneous}, FH-GAT is more like a multi-channel aggregation, which collects field-specific user preferences in categories from multi-hop neighbors for node aggregation.
These aggregated node embeddings are regarded as the final representations for all types of nodes in both offline training and online matching.

\subsection{Diversity-enhanced Training Objective}
\label{sec.objective}

Conventional ranking models usually rely on supervised training with CTR-oriented objectives, which also brings in homogenization. In this work, instead of merely focusing on CTR, GraphDR aims to learn user diverse preferences from multi-aspect factors and improve both accuracy and diversity. Therefore, we conduct the \textbf{neighbor-similarity based loss} \cite{liu2020graph} instead of conventional CTR-oriented objectives to highlight diversity.
Specifically, we assume that \emph{all nodes should be similar to their neighbors} on the diversified preference network regardless of their types. The neighbor-similarity based loss can be viewed as a specialized DeepWalk \cite{perozzi2014deepwalk} with the path length set to be $2$ (too long paths may bring in more noises and computation costs), which is formalized as follows:
\begin{equation}
\begin{split}
J=\sum_{h_k}\sum_{h_i \in N_k}\sum_{h_j \notin N_k}(\log(\sigma({\bm{h}^\top_k}\bm{h}_j))-\log(\sigma({\bm{h}^\top_k}\bm{h}_i))).
\end{split}
\end{equation}
$\bm{h}_k$ is the $k$-th aggregated node embedding given by FH-GAT, and $N_k$ is the neighbor set of the k-th node. $\sigma(\cdot)$ indicates the sigmoid function. We use Adam \cite{kingma2015adam} with negative sampling for training.

The feasibility and necessity of the neighbor-similarity based loss are discussed as follows:
(1) videos that a user may be interested in are very likely to be connected via (multi-step) paths in the diversified preference network. For example, the multi-step path \emph{video:Apple event} $\leftrightarrow$ \emph{tag:iPhone} $\leftrightarrow$ \emph{tag:fast charge} $\leftrightarrow$ \emph{video:new tech of charge} connects two related videos users may watch sequentially. Through the neighbor-similarity based loss, related heterogeneous nodes linked by multi-hop paths in the diversified preference network will have similar representations.
(2) GraphDR focuses on the matching module which values efficiency. Hence, the online multi-channel matching in Sec. \ref{sec:online_serving} conducts an embedding-based retrieval to meet the requirement of efficiency, which ranks videos according to the similarities between different types of embeddings. The neighbor-similarity based loss perfectly matches the embedding-based retrieval for efficient, accurate and diverse matching.

Cooperating with diversified preference network, the neighbor-similarity based loss can well balance both accuracy and diversity, since it calculates video similarities with multiple factors including user watching habit in session, audience community, video content, taxonomy, and content provider.
Precisely, the click-based supervised information used in classical ranking models is collected by two global interactions in GraphDR: video-video edges (for sequential click information in session) and video-user edges (for community-aggregated user-item interactions). These two types of click-based interactions are still the dominating interactions (taking nearly $83\%$ of all interactions in our dataset given in Table \ref{tab:dataset} to ensure the recommendation accuracy. In contrast, the other four interactions related to tags, medias and words mainly provide the generalization ability of node representations to ensure the recommendation diversity.
Comparing with classical CTR-oriented losses that merely focus on clicks, GraphDR jointly considers user diverse preferences from multiple heterogeneous interactions, and thus could achieve better accuracy and diversity in matching.

\section{Online Serving}
\label{sec:online_serving}

We have deployed our GraphDR on the matching module of a widely-used video recommendation system in WeChat Top Stories, which has nearly billion-level daily views generated by million-level users. We introduce the details of online serving.

\subsection{Online Multi-channel Matching}

Online multi-channel matching aims to retrieve hundreds of items from millions of candidates rapidly. GraphDR first builds the user representation with his/her valid watching behaviors $\{\hat{v}_1, \cdots, \hat{v}_m\}$ of videos. To improve the diversity, we conduct a multi-channel matching strategy as in Fig \ref{fig:architecture}, which jointly retrieves video candidates from multiple aspects of representative tags, medias and videos in user historical behaviors.

In the video channel, each video in the valid watching behavior sequence retrieves top $100$ videos with the cosine similarity between two aggregated video embeddings. The weighting score of the i-th video $v_i$ in the video channel is formulated as:
\begin{equation}
\begin{split}
score_i^v=\sum_{j=1}^{m}x_v(ij) \times complete_j \times time_j \times sim(v_i,\hat{v}_j).
\end{split}
\end{equation}
$x_v(ij)$ equals $1$ only if the i-th video $v_i$ is in the top $100$ nearest videos of the j-th video $\hat{v}_j$ in valid watching sequence, and otherwise equals $0$. $complete_j$ is the watching time length percentage of $\hat{v}_j$, which measures the user's satisfaction of $\hat{v}_j$. $sim(v_i,\hat{v}_j)$ represents the cosine similarity calculated by the aggregated node embeddings of $v_i$ and $\hat{v}_j$. We also use $time_j$ to highlight the short-term interests of users as follows:
\begin{equation}
\begin{split}
time_j=\eta \cdot time_{j+1}, \quad time_m=1,
\end{split}
\end{equation}
in which $\eta=0.95$ is a time decay factor.

In the tag and media channels, we first learn user preferences on tags and medias from user historical behaviors. For example, the i-th tag's preference score $p_i^t$ is defined as:
\begin{equation}
\begin{split}
p_i^t=\sum_{j=1}^{m}z_t(ij) \times complete_j \times time_j,
\end{split}
\end{equation}
where $z_t(ij)$ equals $1$ when the i-th tag belongs to $\hat{v}_j$, and otherwise equals $0$. To reduce noises, we only select top $10$ tags $\hat{t}_j$ ranked by $p_i^t$ to form the user preferred tag set $T_u$. Next, each tag in $T_u$ retrieves top $100$ videos with the cosine similarities between tag and video aggregated embeddings. The weighting score of the i-th video in tag channel is calculated as:
\begin{equation}
\begin{split}
score_i^t=\sum_{\hat{t}_j \in T_u} x_t(ij) \times \frac{p^t_j}{\sum_{\hat{t}_k \in T_u}p_k^t} \times sim(v_i,\hat{t}_j).
\end{split}
\end{equation}
$x_t(ij)$ equals $1$ if $v_i$ is in the top $100$ nearest videos of $\hat{t}_j$, and otherwise equals $0$. $sim(v_i,\hat{t}_j)$ indicates the cosine similarity between $v_i$ and $\hat{t}_j$. The weighting score of $v_i$ in media channel $score_i^m$ is calculated similarly as $score_i^t$ of tag channel.

Finally, we combine all three multiple channels in the joint ranking to get the final video weighting scores as follows:
\begin{equation}
\begin{split}
score_i=\lambda_v \cdot score_i^v+ \lambda_t \cdot score_i^t + \lambda_m \cdot score_i^m.
\end{split}
\end{equation}
We rank all videos with their final video weighting scores and select top $500$ videos as the output of GraphDR. We do not use the user group embedding learned by FH-GAT for online matching, since they are coarse-grained user community representations, and user historical behaviors are more informative for individuals. We also abandon the word channel considering the ambiguity in words.

\subsection{Online Deployment}

The online recommendation system mainly contains two modules including ranking and matching. The ranking module adopts ensemble ranking models including DeepFM \cite{guo2017deepfm}, AutoInt \cite{song2019autoint} and AFN \cite{cheng2020adaptive} to model feature interactions between user, item and contexts. Reinforcement learning is also used for long-term and list-wise rewards. In contrast, the matching module aims to retrieve as many appropriate items as possible. Therefore, the matching module consists of dozens of different types of matching strategies from various aspects. Our GraphDR and other compared matching baselines are worked as one of the matching strategies in the matching module. All matching strategies compete with each other, aiming to generate items to be fed into the the same shared ranking module. Sec. \ref{sec.online_evaluation} gives the implementation details of an online matching evaluation.

Online matching especially values efficiency. In GraphDR, all embedding similarities like $sim(v_i,\hat{v}_j)$ are pre-calculated in offline, which enables fast retrieval. Its online time complexity is less than $O(\log n)$ w.r.t the corpus size $n$, which is much superior to most deep ranking models involving complicated user-item interactions.

\section{Experiments}

In experiments, we conduct extensive offline and online evaluations with detailed analyses on a real-world recommendation system to verify that GraphDR can improve both accuracy and diversity.
In this section, we attempt to answer the following five research questions:
(\textbf{RQ1}): How does the proposed GraphDR model perform against different types of competitive models on recommendation accuracy in matching (see Sec. \ref{sec.accuracy})?
(\textbf{RQ2}): How does GraphDR perform against competitive baselines on recommendation diversity at element level, list level and global level (see Sec. \ref{sec.diversity})?
(\textbf{RQ3}): How does PAPERec perform in online system with various online accuracy and diversity related evaluation metrics (see Sec. \ref{sec.online_evaluation})?
(\textbf{RQ4}): How do different essential parameters affect GraphDR on recommendation accuracy and diversity (see Sec. \ref{sec.model_analyses})?
(\textbf{RQ5}): Will node representations learned by GraphDR be successfully encoded with user diverse preferences (see Sec. \ref{sec.case_study})?

\subsection{Datasets}

Since there are few large-scale datasets for evaluating recommendation accuracy and diversity in matching, we build a novel dataset DivMat-2.1B extracted from WeChat Top Stories. We randomly select nearly $15$ million users, collect their $2.1$ billion video watching instances after data masking for privacy, and split the dataset into a train set and a test set using the chronological order. In train set, we build a huge diversified preference network following Sec. 3.2, where $15$ million users are aggregated into $93$ thousand user groups (users in the same user group have the same gender-age-location attribute triplet).  The test set contains $8,132,719$ valid watching behavior instances for offline evaluation in matching.

\begin{table}[!hbtp]
\centering
\small
\caption{Statistics of the DivMat-2.1B dataset.}
\begin{tabular}{p{0.9cm}<{\centering}|p{0.85cm}<{\centering}|p{0.85cm}<{\centering}|p{0.95cm}<{\centering}|p{0.9cm}<{\centering}|p{1.08cm}<{\centering}}
\toprule
video & user & tag & media & word & instance \\
\midrule
1.2M & 15M & 103K & 74K & 150K & 2.1B \\
\bottomrule
\end{tabular}
\begin{tabular}{p{0.92cm}<{\centering}|p{0.92cm}<{\centering}|p{0.92cm}<{\centering}|p{0.92cm}<{\centering}|p{0.92cm}<{\centering}|p{0.92cm}<{\centering}}
\toprule
\#v-v & \#v-t & \#v-m & \#v-w & \#v-u & \#t-t \\
\midrule
97M & 6.1M & 1.2M & 8.1M & 2.3M & 5.3M \\
\bottomrule
\end{tabular}
\label{tab:dataset}
\end{table}

\subsection{Competitors}
\label{sec.competitor}

We implement several classical models as baselines, and categorize these competitors into four groups.

\textbf{\emph{IR-based Methods.}}
We implement three IR-based methods including Category-based, Tag-based and Media-based IR methods \cite{khribi2008automatic}. For Tag-based method, we build a tag-video inverted index, where videos are ranked with their popularity. The online matching retrieves videos with user preferred tags. Other IR-based methods are similar to Tag-based IR method.

\textbf{\emph{CF-based Methods.}}
We implement Item-CF \cite{sarwar2001item} to retrieve similar videos with video co-occurrence. Moreover, we also implement BERT-CF, which uses semantic similarity to measure video similarity. Precisely, we calculate the semantic similarity of two videos with their title embeddings learned by BERT \cite{devlin2019bert}, and conduct CF to learn video embeddings for fast retrieval.

\textbf{\emph{Homogeneous NRL Methods.}}
We implement some typical NRL models on the homogeneous video network built with video sessions. The compared methods include DeepWalk \cite{perozzi2014deepwalk} and GraphSAGE \cite{hamilton2017inductive}. These learned video representations are then used for online embedding-based matching with the video channel.

\textbf{\emph{Neural-based Methods.}}
Youtube candidate generation model \cite{covington2016deep} is a classical deep model for matching. We further improve the original Youtube model with behavior-level attention \cite{zhou2018deep} and neural FM \cite{he2017neural} as Youtube+ATT+FM, which is a strong industrial baseline in practice. Moreover, we implement DSSM \cite{huang2013learning}, which retrieves items according to the user-item similarities. We also implement AutoInt \cite{song2019autoint} to model feature interactions. These models are optimized under supervised learning with video behaviors.

We conduct a nearest neighbor server for all embedding-based fast retrieval.
Note that we do not compare with complicated diversified recommendation models specially designed for ranking, due to their tremendous computation costs in matching \cite{karakaya2018effective}. We do not report TDM/JDM either for the static tree-based retrieval is challenging to handle various aspects of diversities in videos.

\textbf{\emph{Ablation Test Settings.}} We implement the heterogeneous versions of GraphSAGE \cite{hamilton2017inductive} and GAT \cite{velivckovic2018graph} to replace FH-GAT in the NRL module for ablation tests. We use GraphDR(GraphSAGE) and GraphDR(GAT) to represent these two settings respectively.

\subsection{Experimental Settings}

In GraphDR, the node feature embedding dimension is $900$, where the video field's dimension $d_v$ is $300$ and others' are $150$. The dimensions of two output embeddings in FH-GAT are $120$. The numbers of neighbor sampling in the first and second layers are $30$ and $20$. In training, we randomly select $20$ negative samples for each positive sample, and set batch size as $512$. In online matching, we consider top $200$ recent watched videos and retrieve top $500$ candidates for ranking.
The weighting scores $\lambda_v$, $\lambda_t$ and $\lambda_m$ are equally set to be $1$. We conduct the grid search for parameter selection. For fair comparisons, all models follow the same settings in evaluation.

\subsection{Recommendation Accuracy (RQ1)}
\label{sec.accuracy}

We first evaluate all GraphDR models and baselines on recommendation accuracy in offline DivMat-2.1B dataset.

\subsubsection{Evaluation Protocols}

We focus on matching that aims to generate \textbf{hundreds of} item candidates. Differing from ranking, matching only cares \textbf{whether good items are retrieved}, not the specific item ranks. Therefore, we use hit rate (HIT@N) \cite{sun2019bert4rec} as the evaluation metric for accuracy, where an instance is ``hit'' if the clicked item is ranked in top N.
We do not use classical ranking metrics such as MAP and NDCG since matching does not care specific ranks.
To simulate the real-world scenarios, we conduct HIT@N with N set as $100$, $200$, $300$ and $500$. Since we retrieve \textbf{top} $\textbf{500}$ items in the online recommendation system, HIT@$500$ is considered to be the most essential accuracy metric.

\begin{table}[!hbtp]
\centering
\small
\caption{Results of recommendation accuracy. We set N=500 in the matching module of our online system.}
\begin{tabular}{l|c|c|c|c}
\toprule
HIT@N & N=100 & N=200 & N=300 & N=500 \\
\midrule
Category-based & 0.0010 & 0.0018 & 0.0021 & 0.0031 \\
Tag-based & 0.0157 & 0.0207 & 0.0240 & 0.0287 \\
Media-based & 0.0235 & 0.0297 & 0.0337 & 0.0383 \\
\midrule
BERT-CF & 0.0337 & 0.0469 & 0.0556 & 0.0669 \\
Item-CF & 0.0748 & 0.0904 & 0.1214 & 0.1459 \\
\midrule
DeepWalk & 0.0799 & 0.0998 & 0.1130 & 0.1340 \\
GraphSAGE & 0.0932 & 0.1242 & 0.1568 & 0.1862 \\
\midrule
DSSM & 0.1012 & 0.1326 & 0.1631 & 0.2031 \\
AutoInt & 0.1087 & 0.1488 & 0.1892 & 0.2401 \\
Youtube+ATT+FM & \textbf{0.1392} & \textbf{0.1892} & \underline{0.2194} & 0.2549 \\
\midrule
GraphDR(GraphSAGE) & 0.1013 & 0.1442 & 0.1818 & 0.2372 \\
GraphDR(GAT) & 0.1088 & 0.1674 & 0.2108 & \underline{0.2731} \\
GraphDR(FH-GAT) & \underline{0.1241} & \underline{0.1885} & \textbf{0.2384} & \textbf{0.3102} \\
\bottomrule
\end{tabular}
\label{tab:accuracy}
\end{table}

\begin{table*}[!hbtp]
\caption{Results of different evaluation metrics on recommendation diversity.}
\centering
\small
\begin{tabular}{l||p{1.1cm}<{\centering}|p{1.1cm}<{\centering}|p{1.1cm}<{\centering}||
p{1.1cm}<{\centering}|p{1.1cm}<{\centering}|p{1.1cm}<{\centering}||
p{1.1cm}<{\centering}|p{1.1cm}<{\centering}|p{1.1cm}<{\centering}}
\toprule
\multirow{2}{*}{Model} & \multicolumn{3}{c||}{Element-level diversity} & \multicolumn{3}{c||}{List-level diversity} & \multicolumn{3}{c}{Global-level diversity} \\
\cmidrule{2-10}
 & tag & cate & media & tag & cate & media & coverage & long-tail & novelty \\
\midrule
Category-based & 17.64 & 1.00 & 13.15 & 206.98 & 4.26 & 98.76 & 0.0012 & 0.0836 & 0.0043 \\
Tag-based & 24.39 & 1.91 & 12.20 & 346.42 & 23.31 & 315.48 & 0.0270 & 0.1432 & 0.0343 \\
Media-based & 29.67 & 2.95 & 1.00 & 434.30 & 43.41 & 9.58 & 0.0309 & 0.1327 & 0.0543 \\
\midrule
BERT-CF & 26.29 & 2.27 & 11.06 & 387.45 & 30.52 & 207.41 & 0.3829 & 0.2631 & 0.5734 \\
Item-CF & 31.86 & 3.66 & 11.47 & 499.42 & 55.42 & 234.31 & 0.1786 & 0.0000  & 0.3143 \\
\midrule
DeepWalk & 30.64 & 3.24 & 13.23 & 476.76 & 52.53 & 246.33 & 0.1642 & 0.0000 & 0.3821 \\
GraphSAGE & 31.67 & 2.84 & 13.65 & 426.32 & 41.11 & 285.52 & 0.1806 & 0.0000 & 0.3532 \\
\midrule
DSSM & 25.15 & 2.13 & 13.94 & 363.41 & 29.65 & 211.32 & 0.1688 & 0.0525 & 0.2843 \\
AutoInt & 26.31 & 2.41 & 13.21 & 372.31 & 32.12 & 242.31 & 0.1762 & 0.0612 & 0.2971 \\
Youtube+ATT+FM & 31.22 & 2.79 & 12.83 & 457.15 & 41.93 & 217.67 & 0.1532 & 0.0734 & 0.3523 \\
\midrule
GraphDR(GraphSAGE) & 33.19 & 3.61 & 14.91 & 498.31 & 51.21 & 327.28 & 0.4892 & 0.2854 & 0.6742 \\
GraphDR(GAT) & \underline{34.77} & \underline{3.79} & \underline{15.34} & \underline{516.93} & \underline{56.62} & \underline{358.82} & \underline{0.4934} & \underline{0.3242} & \underline{0.7032} \\
GraphDR(FH-GAT) & \textbf{37.15} & \textbf{3.96} & \textbf{16.43} & \textbf{538.32} & \textbf{63.41} & \textbf{379.12} & \textbf{0.5132} & \textbf{0.3678} & \textbf{0.7352} \\
\bottomrule
\end{tabular}
\label{tab:diversity}
\end{table*}

\subsubsection{Experimental Results}

In Table \ref{tab:accuracy} we can observe that:

(1) GraphDR(FH-GAT) significantly outperforms all baselines on HIT@$500$ with the significance level $\alpha=0.01$. It indicates that GraphDR(FH-GAT) could retrieve accurate items in matching. Differing from conventional CTR-oriented models, GraphDR considers user diverse preferences related to video session, community, taxonomy, semantics and provider, which makes the matching results more diversified. GraphDR is perfectly suitable for matching, since it concerns more about item coverage than their specific ranks.

(2) GraphDR(FH-GAT) performs comparable or slightly worse than Youtube+ATT+FM when $N$ is small. It is intuitive since the neighbor-similarity based loss should balance accuracy and diversity, which inevitably harms ranking accuracy (not matching). In contrast, Youtube is a strong supervised baseline that benefits from its CTR-oriented objective. However, it suffers from overfitting and homogenization, and thus performs much worse than GraphDR(FH-GAT) when N grows bigger (which is the practical scenario).
The diversity issue will be discussed in Sec. \ref{sec.diversity}.

(3) Both IR-based methods and BERT-CF are not satisfactory. It indicates that the taxonomy and semantic similarities contribute less to accuracy compared to user behaviors. In contrast, Neural-based methods focus on CTR-oriented objectives and thus get better accuracies. However, they still perform worse than GraphDR, for they fail to consider heterogeneous interactions and thus lack coverage.

\textbf{Ablation study.} Among different GraphDR versions, we find that FH-GAT outperforms GAT and GraphSAGE. It confirms the power of field-specific aggregation in modeling user diverse preferences. Moreover, we further conduct an ablation test to verify that all different types of nodes are necessary for the diversified recommendation. For instance, the HIT@$500$ will drop to $29.31\%$ if we wipe out all word nodes in DivMat-2.1B.

\subsection{Recommendation Diversity (RQ2)}
\label{sec.diversity}

In this subsection, we evaluate all models on both individual diversity and aggregate diversity in recommendation with various evaluation metrics.

\subsubsection{Evaluation Protocols}

We conduct nine typical diversity metrics and group them into three classes, namely the element-level diversity, the list-level diversity and the global-level diversity. The former two diversities indicate the individual diversity, while the latter diversity measures the aggregate diversity \cite{kunaver2017diversity}.
The \textbf{element-level diversity} focuses on the diversity in each element, such as the tag, category, media in IR-based methods and the embeddings in baselines. Precisely, we regard the average deduplicated tag/category/media numbers in top $20$ videos retrieved by these elements as the element-level diversity.
The \textbf{list-level diversity} measures diversity in recommended lists (top $500$ items).
We use the average deduplicated tag/category/media numbers in the final recommended lists as the list-level diversity \cite{ziegler2005improving,wu2016relevance}.
In the \textbf{global-level diversity}, \emph{coverage} indicates the percentage of items that could be recommended \cite{karakaya2018effective}. \emph{Long-tail} indicates the percentage of long-tail items in all results (videos that have not been watched for $15$ days are empirically viewed as the long-tail videos). \emph{Novelty} represents the percentage of new items generated by this model that other models do not recommend \cite{zhang2008avoiding}.

\begin{table*}[!hbtp]
\caption{Online A/B test on recommendation accuracy and diversity in a real-world system.}
\centering
\small
\begin{tabular}{p{3.2cm}||p{1.35cm}<{\centering}|p{1.35cm}<{\centering}|p{1.35cm}<{\centering}|p{1.35cm}<{\centering}|p{1.35cm}<{\centering}|p{1.4cm}<{\centering}|p{1.45cm}<{\centering}}
\toprule
 & VV & VWT/c & VWT/v & PT & DIV & Tag diver & Cate diver \\
\midrule
GraphSAGE & +3.08\% & +6.20\% & +1.48\% & +4.66\% & +2.43\% & +6.24\% & +10.27\% \\
GraphDR(GraphSAGE) & +4.37\% & +7.61\% & +1.68\% & +6.04\% & +3.97\% & +9.16\% & +12.42\% \\
GraphDR(GAT) & +5.30\% & +9.36\% & +2.49\% & +6.07\% & +8.00\% & +12.81\% & +15.57\% \\
GraphDR(FH-GAT) & \textbf{+6.08\%} & \textbf{+10.79\%} & \textbf{+3.10\%} & \textbf{+6.10\%} & \textbf{+10.43\%} & \textbf{+14.68\%} & \textbf{+17.00\%} \\
\bottomrule
\end{tabular}
\label{tab:online}
\end{table*}

\subsubsection{Experimental Results}

Table \ref{tab:diversity} shows the results of various diversity metrics, form which we can know that:

(1) GraphDR(FH-GAT) achieves the best performances in all diversity metrics. The improvement derives from all three modules: (i) in diversified preference network, the heterogeneous interactions store user diverse preferences on taxonomy, semantics, community, video session and provider to link similar videos via multi-hop paths. (ii) In NRL, FH-GAT and its neighbor-similarity based loss successfully encode user diverse preferences into node representations. (iii) In online matching, the multi-channel strategy retrieves items from tag/media/video aspects, which also amplifies diversity.
In addition, GraphDR(GraphSAGE) and GraphDR(GAT) generally outperform all baselines but still inferior to GraphDR(FH-GAT). It reconfirms the power of FH-GAT in diversity.

(2) The element-level and list-level diversities indirectly measure the individual diversity with diversities in tag, category and media. We assume that more tags/medias/categories in recommended lists indicate a more diversified recommendation. We find that behavior-based models like Youtube and GraphSAGE perform better than other baselines in individual diversities. Nevertheless, GraphDR has better results since it considers other types of interactions.

(3) The global-level diversity measures the aggregate diversity, where coverage, long-tail and novelty focus on different aspects.
Behavior-based models only consider video watching behaviors, which are hard to handle long-tail and new items.
In contrast, BERT-CF focuses on content similarity and achieves good aggregate diversity. Still, GraphDR considers user diverse preferences in various fields and achieves the best aggregate diversity.

\subsection{Online Evaluation (RQ3)}
\label{sec.online_evaluation}

The offline evaluation has verified the improvements of accuracy and diversity in matching module.
We further conduct an online A/B test to evaluate GraphDR in real-world industrial-level scenarios.

\subsubsection{Evaluation Protocols}

We implement GraphDR on the matching module of WeChat Top Stories following Sec. \ref{sec:online_serving}.
The original online matching model is an ensemble model containing multiple IR-based, CF-based and Neural-based methods in Sec. \ref{sec.competitor}. We regard GraphDR as an additional matching channel to the existing online ensemble model, with the ranking module unchanged. All videos retrieved by different matching channels will jointly compete with each other in the following ranking module.

In online A/B test, we focus on the following seven representative metrics to evaluate accuracy and diversity:
(1) video views per capita (VV), (2) video watching time per capita (VWT/c), (3) video watching time per video (VWT/v), (4) page turns per capita (PT), (5) deduplicated impressed videos per capita (DIV), (6) watched tag per capita (Tag diver), and (7) watched category per capita (Cate diver). The former five metrics mainly measure accuracy, while the latter two measure diversity.
We conduct the A/B test for $5$ days with nearly $3.8$ million users involved, and report the improvement percentages over the ensemble base model. The online evaluation can be viewed as an online ablation test.

\subsubsection{Experimental Results}

Table \ref{tab:online} shows the results of online evaluation with multiple metrics, from which we find that:

(1) All GraphDR models outperform the ensemble base model, among which GraphDR(FH-GAT) achieves the best performances in accuracy and diversity with the significance level $\alpha=0.01$. We have also passed the homogeneity test in online evaluation, which confirms that the system and traffic split are unbiased and the improvements are stable. It verifies the effectiveness of GraphDR in real-world scenarios. Moreover, the improvements from GraphSAGE to FH-GAT also imply the significances of FH-GAT.

(2) The significant improvements in the former five metrics reflect better accuracy. A better video view metric indicates that users are more willing to click videos, while a better video watching time indicates users are genuinely interested in their clicked videos. Moreover, the page turns and deduplicated impressed video metrics also reflect user experiences indirectly. Users will slide down and browse more videos if they are satisfied with the results.

(3) The average watched tags and categories measure the diversity. The better tag/category diversity derives from two factors: more diverse videos impressed to users, and better personalized results that attract users to watch more videos. These diverse items help us to explore users' potential interests and give surprising results, which could even contribute to the long-term performances.

\subsection{Model Analyses (RQ4)}
\label{sec.model_analyses}

We conduct several analyses on different channels and user behavior sequence lengths to better understand GraphDR.

\subsubsection{Analysis on Multi-channel Matching}

In GraphDR, the online multi-channel matching module plays an important role in improving diversity.
We evaluate the GraphDR(FH-GAT) on HIT@N and list-level diversity metrics with different channels individually. From Table \ref{tab:multi_channel} we find that: the video channel achieves better HIT@N results, since video embeddings are directly influenced by video watching behaviors. In contrast, the tag and media channels are more responsible for diversity. To balance both accuracy and diversity, we combine all three channels in GraphDR.

\begin{table}[!hbtp]
\caption{Results of different matching channels.}
\centering
\small
\begin{tabular}{p{2.0cm}<{\centering}|p{1.0cm}<{\centering}|p{1.0cm}<{\centering}|p{1.0cm}<{\centering}|p{1.0cm}<{\centering}}
\toprule
Channel & tag & media & video & joint \\
\midrule
 HIT@100 & 0.1027 & 0.0934 & \textbf{0.1323} & 0.1241 \\
 HIT@200 & 0.1571 & 0.1497 & \textbf{0.1943} & 0.1885 \\
 HIT@300 & 0.2143 & 0.2032 & \textbf{0.2512} & 0.2384 \\
 HIT@500 & 0.2787 & 0.2583 & \textbf{0.3312} & 0.3102 \\
 \midrule
 Tag diversity & \textbf{573.43} & 543.31 & 468.42 & 538.32 \\
 Cate diversity & \textbf{71.31} & 68.32 & 53.63 & 63.41 \\
 Media diversity & 387.48 & \textbf{401.58} & 344.32 & 379.12 \\
\bottomrule
\end{tabular}
\label{tab:multi_channel}
\end{table}

\subsubsection{Analysis on Behavior Sequence Length}

We also analyze the impacts of different behavior sequence lengths in online matching. In Table \ref{tab:behavior_length}, as the behavior sequence length increases, HIT@N metrics achieve consistent improvements, while diversity metrics become slightly worse. It implies that considering user long-term preferences can better understand users in recommendation. However, user long-term preferences are more stable, which inevitably harm the diversity. In GraphDR, we set the length as $200$ since the improvements in accuracy are more significant than diversity.

\begin{table}[!hbtp]
\caption{Results of different user behavior lengths.}
\centering
\small
\begin{tabular}{p{2.0cm}<{\centering}|p{1.0cm}<{\centering}|p{1.0cm}<{\centering}|p{1.0cm}<{\centering}|p{1.0cm}<{\centering}}
\toprule
Length & m=20 & m=50 & m=100 & m=200 \\
\midrule
 HIT@100 & 0.0791 & 0.0883 & 0.1072 & \textbf{0.1241} \\
 HIT@200 & 0.1237 & 0.1373 & 0.1653 & \textbf{0.1885} \\
 HIT@300 & 0.1742 & 0.1902 & 0.2114 & \textbf{0.2384} \\
 HIT@500 & 0.2393 & 0.2617 & 0.2763 & \textbf{0.3102} \\
 \midrule
 Tag diversity & \textbf{556.12} & 552.22 & 547.43 & 538.32 \\
 Cate diversity & \textbf{69.52} & 68.11 & 66.73 & 63.41 \\
 Media diversity & \textbf{395.45} & 391.52 & 387.91 & 379.12 \\
\bottomrule
\end{tabular}
\label{tab:behavior_length}
\end{table}

\subsection{Case Study (RQ5)}
\label{sec.case_study}

In GraphDR, user diverse preferences are encoded in node embeddings. We give some tags and their nearest tags to explicitly display the diversity in Table \ref{tab:tag_nearest}.
The interest in \emph{Restaurant guide} may expand to specific food like \emph{Foie gras} and their stories like \emph{Food documentary}. The nearest tags of \emph{El Nino phenomenon} reflect the interests in nature and science. Users like \emph{iPhone 11 Pro Max} may also seek information on its hardware, software, and discount information. These nearest tags reflect both similarities in semantics and user preferences, since the node representations are learned under the neighbor-similarity based objective with a diversified preference graph containing various heterogeneous feature interactions. Similar phenomenon can be found in other nodes.

\begin{table}[!hbtp]
\caption{Examples of tags and their nearest tags.}
\centering
\small
\begin{tabular}{p{1.7cm}<{\centering}|p{5.6cm}}
\toprule
Tag & Nearest tags \\
\midrule
Restaurant guide & Roasted goose; Food documentary; Melaleuca cake; Foie gras; Hong Kong cuisine \\
\midrule
El Nino phenomenon & Superluminal speed; Easter island; Darwin; Absolute zero; Parallel worlds theory \\
\midrule
iPhone 11 Pro Max & iPhone SE; Fast charge; Mobile phone test; Voice assistant; iPhone discount \\
\bottomrule
\end{tabular}
\label{tab:tag_nearest}
\end{table}

Table \ref{tab:user_nearest} shows the nearest tags of some typical user groups. According to the node embeddings and aggregated behaviors, young men users in our dataset are more interested in sports, while young women focus more on fashion. Differing from the youth, the elderly in Beijing concentrate on traditional Chinese art and culture. The geographic distance also leads to fine-grained differences in interested sports (e.g., golf V.S. soccer). The preference divergences in different communities verify the success of diversity modeling.

\begin{table}[!hbtp]
\caption{Examples of user groups with nearest tags.}
\centering
\small
\begin{tabular}{c|c|c|p{5.0cm}}
\toprule
Sex & Age & City & Nearest tags \\
\midrule
M & 21 & Beijing & Sports news; Entrepreneur; Comedy; Scientific anecdotes; Soccer \\
\midrule
F & 21 & Beijing & Summer wear; Constellation; Product promotion; Diet food; Potted plant \\
\midrule
M & 59 & Beijing & Calligraphy; Social documentary; Tai Chi; Exercise; Family \\
\midrule
M & 21 & London & London Olympics; The Celtic; Scientists; Golf; 100 metres race\\
\bottomrule
\end{tabular}
\label{tab:user_nearest}
\end{table}

\section{Conclusion and Future Work}

In this work, we propose a simple and effective GraphDR framework to improve both accuracy and diversity in matching. We propose a new diversified preference network to capture heterogeneous interactions between essential objects in recommendation. We also design a novel FH-GAT model with a neighbor-similarity based loss to encode user diverse preferences from heterogeneous interactions. In experiments, we conduct extensive offline and online evaluations, model analyses and case studies. The significant improvements verify the effectiveness and robustness of GraphDR in improving accuracy and diversity simultaneously.

In the future, we will explore more types of interactions and weighted edges in GraphDR. Moreover, we will enhance the multi-channel matching with more sophisticated models. Better graph neural networks are also worth being studied, which will be easily adopted in our GraphDR framework.

\bibliographystyle{ACM-Reference-Format}
\bibliography{reference}

\end{document}